\documentclass{iau}
\usepackage{graphicx}

\def\la{\raise.5ex\hbox{$<$}\kern-.8em\lower 1mm\hbox{$\sim$}}
\def\ga{\raise.5ex\hbox{$>$}\kern-.8em\lower 1mm\hbox{$\sim$}}
\def\be{\begin{equation}}
\def\ee{\end{equation}}
\def\ba{\begin{eqnarray}}
\def\ea{\end{eqnarray}}

\def\Mdot*{\dot{M}_*}
\def\Mdotin{\dot{M}_{\mathrm{in}}}
\def\Mdot{\dot{M}}

\def\dM*{\delta M_*}

\def\P0min{P_{0,{\mathrm{min}}}}

\title[Fallback Disks] 
{Fallback Disks, Magnetars and Other Neutron Stars}

\author[Alpar et al.]   
{M. Ali Alpar, \c{S}. \c{C}al\i\c{s}kan \and \"{U}. Ertan}

\affiliation{Sabanc\i \,University, \.Istanbul, Turkey  \\ email: {\tt alpar@sabanciuniv.edu} \\[\affilskip]}

\pubyear{2012}
\volume{290}  
\jname{Feeding compact objects: Accretion on all scales}
\editors{C.M. Zhang, T. Belloni, M. M\'endez \& S.N. Zhang, eds.}

\begin{document}

\maketitle

\begin{abstract}
The presence of matter with angular momentum, in the form of a fallback disk around a young isolated neutron star will determine its evolution. This leads to an understanding  of many properties of different classes of young neutron stars, in particular a natural explanation for the period clustering of AXPs, SGRs and XDINs. The spindown or spinup properties of a neutron star are determined by the dipole component of the magnetic field. The natural possibility that magnetars and other neutron stars may have different strengths of the dipole and higher multipole components of the magnetic field is now actually required by observations on the spindown rates of some magnetars. This talk gives a broad overview and some applications of the fallback disk model to particular neutron stars.  Salient points are: (i) A fallback disk has already been observed around the AXP 4U 0142+61 some years ago. (ii) The low observed spindown rate of the SGR 0418+5729 provides direct evidence that the dipole component of the field is in the $10^{12} G$ range. All properties of the SGR 0418+5729 at its present age can be explained by spindown under torques from a fallback disk.  (iii) The anomalous braking index of PSR J1734-3333 can also be explained by the fallback disk model which gives the luminosity, period, period derivative {\em and} the period second derivative at the present age. (iv) These and all applications to a variety of other sources employ the same disk physics and evolution, differing only in the initial conditions of the disk.   

\end{abstract}

\firstsection 
\section{Introduction}

There are many different classes of  neutron stars occupying various parts of the $P\dot{P}$ diagram: Compact Central Objects (CCOs),  Anomalous X-ray Pulsars (AXPs) \& Soft Gamma-Ray Repeaters (SGRs), X-Ray Dim Isolated Neutron Stars (XDINs),  Rotation Powered  (mostly radio) Pulsars  - including RRATs , intermittents,  accreting  X-ray  millisecond  Pulsars and millisecond  Radio Pulsars. How are these different classes of neutron stars related to each other? Are there evolutionary connections between the classes of young neutron stars as established for old populations, the low mass X-ray binaries, accreting  X-ray  millisecond  pulsars and millisecond  Radio Pulsars?  Are the various categories of young neutron stars subclasses of a single initial distribution? Is there a ``birthrate problem" in matching the inferred birthrates of young neutron stars with the inferred supernova rate? How do we explain the locations on the $P\dot{P}$ diagram and other properties of the different classes of neutron stars? How do we understand defining properties, like period clustering, that is common to AXPs, SGRs and XDINs, and variations within each class like the braking indices of magnetars and radio pulsars? 

The spindown or spinup properties of a neutron star are determined by the dipole component of the magnetic field.  If  the torque on the neutron star is the pure dipole radiation torque in vacuum, the observed period $P$ and the period derivative $\dot{P}$ give the dipole magnetic moment $\mu_{dipole}$ through the relation
\begin{equation}
P\dot{P} = \frac{8 \pi^2}{3 \; I c^3} \mu_{dipole}^2,
\end{equation}   
where $I \sim 10^{45} gm\, cm^{2}$ is the moment of inertia of the neutron star and $c$ is the speed of light. The dipole magnetic moment is related to the dipole magnetic field $B_{dipole}$ on the neutron star surface, at the magnetic equator, through $\mu_{dipole} =  B_{dipole} R^3$ where $R$ is the neutron star radius. With torques arising from other mechanisms of interaction between the neutron star and its surroundings, the observed $P$ and $\dot{P}$ will lead to different estimates for the dipole magnetic moment $\mu_{dipole}$. 

The bursts  of AXPs and SGRs are believed to be powered by very strong ``magnetar" fields on the neutron star surface. These magnetars, with $P \sim 2-12 s$ and large $\dot{P}$ share the upper right hand corner of the $P\dot{P}$ diagram with radio pulsars with strong dipole magnetic moments 
(\cite[Woods \& Thompson 2006]{WT06}, 
\cite[Mereghetti 2008]{Mer08},
\cite[G\"{o}\u{g}\"{u}\c{s} 2011]{Gogus11}). There is an interesting crossover in the behaviour of AXPs and SGRs and the strong field radio pulsars. The transient  AXP XTE  J1810--197 has X-ray outbursts and radio pulsar emission (\cite[Ertan \& Erkut 2008]{ErtanErkut08}) and two other magnetars show occasional radio pulsar behaviour (\cite[Camilo et al. 2007]{Camiloetal07},  \cite[Levin et al. 2010]{Levinetal10}).
There are strong field radio pulsars  - with inferred dipole fields of $\sim 5 \times 10^{13}\; G$ (\cite[Ng \& Kaspi 2011]{NgKaspi11}), comparable to the inferred dipole fields of many magnetars, and transients that  act as radio pulsars with inferred magnetar fields like 1E 1547.0--5408, which has no X-ray pulses, and brightens by a factor of 16 in X-rays in going from its quiescent phase to its X-ray luminous phase. Radio pulses at a period $P = 2 s$ are detected from this source (\cite[Camilo et al. 2007]{Camiloetal07}). Its inferred dipole field $B_{dipole} = 2.2 \times 10^{14}\, G$  is in the magnetar range. On the other hand there are magnetars with inferred dipole fields less than many typical (not even high field) rotation powered pulsars, like SGR 0418 + 5729 (\cite[Rea et al. 2010]{Reaetal10}), which we discuss in more detail in this talk, and SWIFT J1822.3-1606 (\cite[Rea et al. 2012a]{Reaetal12a}). All of this suggests that the surface dipole field may be different from, possibly much smaller than, the total surface field that is responsible for the magnetar bursts, and that in some of the sources, under torques involving matter around the neutron star, the dipole field may be different from what is inferred from the vacuum dipole radiation formula (Eq. 1.1). 

If dipole spindown into vacuum is taken to be the dominant mechanism governing the evolution of the neutron star, then there are two initial conditions on a newborn neutron star that will determine its subsequent evolution: the initial rotation rate ${\bf \Omega}_0$ (initial  period $P_0$) and the dipole magnetic moment ${\bf \mu}_{dipole}$. If the dipole moment's component $\mu_{dipole, \perp}$ perpendicular to ${\bf \Omega}_0$  remains constant, the pulsar will evolve on lines of slope -1 in the $P\dot{P}$ diagram. Dipole spindown into vacuum does not explain the $P\dot{P}$ diagram:  Not only the different populations and the distribution of sources in the $P\dot{P}$ diagram but also properties of the rotation powered pulsars, in particular braking indices not equal to 3 require new physics. This is likely to involve matter around the neutron star. In order to have evolutionary effects, sustained presence of matter bound to the neutron star is needed. Such bound matter should have angular momentum.  The limited range of AXP, SGR and XDIN periods, $P \sim 2-12 s$ (period clustering) also points to a store of angular momentum that regulates the neutron star's rotation rate -- a gyrostat. A fallback disk left over from the collapse of the progenitor’s core in the supernova explosion was proposed by  Chatterjee, Hernquist \& Narayan (2000) to explain the AXPs. Alpar (2001) proposed that the presence or absence and properties of fallback disks is the third initial condition on newborn neutron stars, in addition to ${\bf \mu}_{dipole}$ and $P_0$: this third parameter determines the different paths of subsequent evolution leading to the different classes of young neutron stars observed.  A fallback disk has a limited active lifetime, unlike disks in binaries, whose active lifetime can extend to evolutionary timescales of the binary and the companion, as long as the mass transfer from the companion continues. The evolution of the fallback disk in interaction with the neutron star leads to the terminal conditions for the observably luminous phase of the source, in particular, to the observed period clustering.

\section{Evolution of a Neutron Star with a Fallback Disk}

Model evolution scenarios for a neutron star with a fallback disk start with choosing $P_0$ and $\mu_{dipole, \perp}$ from distributions inferred by population synthesis for the radio pulsars (\cite[e.g. Faucher-Giguere \& Kaspi 2006]{F-GK06}) and choosing an initial mass and angular momentum for the fallback disk. At each stage the light cylinder radius $r_{lc} = c\,P/(2\pi)$ and the inner radius $r_{in}  = r_{A} = \Mdotin^{-2/7}(G M_{\ast })^{-1/7}\mu _{\ast }^{4/7}     $ are calculated. The inner radius of the disk $r_{in}$, taken as the Alfv$\acute{e}$n radius $r_{A}$, is determined by the current mass inflow rate through the disk, $\dot{M}(t)$. When the inner disk protrudes the light cylinder the mass inflow or a fraction of it will be accreted onto the neutron star, giving an X-ray luminosity due to accretion onto the star. Accretion quenches radio emission which is generated at an inner gap close to the neutron star. The spindown torque applied by the disk determines the rotational evolution of the pulsar. The disk torque $N_{disk}$ depends on the fastness parameter $\omega \equiv \Omega_{*}/\Omega_K(r_A)$. Taking this dependence to be $N_{disk} \propto (1 - \omega^2)$ gives consistent results in model applications for all different classes of neutron stars. In later epochs the disk inner radius can recede back to the light cylinder, after which it tracks the light cylinder and torques are calculated with $r_{in} \cong r_{lc}$. The accretion luminosity then drops. The X-ray luminosity in such epochs is just the cooling luminosity from the neutron star surface, which
can be $L_x \sim 10^{33}\, erg~s^{-1}$ for the first $\sim10^6\, yrs$ of the neutron star's life, dropping sharply to even lower luminosities after that. 

The irradiation of the disk by the X-rays from the neutron star is taken into account in the  evolution, with an irradiation parameter $C \sim 10^{-4}$. During this phase, the pulsar magnetosphere can be active in the presence of the disk within the magnetosphere (\cite[Michel \& Dessler 1981]{MichelDessler81}, \cite[Cheng \& Ruderman 1991]{CR91}), giving rise to optical and infrared pulsar activity from a disk dynamo porcess (\cite[Ertan \& Cheng 2004]{ErtanCheng04}). In some epochs, relevant for an understanding of transient AXPs and the occasional presence of radio pulsar emission, the disk may go back and forth between hot and cold states, characterized by viscosity parameters $\alpha_{hot} \cong 0.1$ and $\alpha_{cold} \cong 0.03$ respectively (\cite[\c{C}al\i\c{s}kan \& Ertan 2012]{CaliskanErtan12}). 

As a fallback disk evolves the mass inflow rate $\dot{M}(t)$ decays, leading to decreasing energy dissipation rates and temperature profiles.  Starting from the outer (colder) disk regions, temperatures will eventually fall below a critical level $T_p$ corresponding to the lowest ionization fraction that can generate viscosity in the disk, presumably through the Balbus-Hawley magneto-rotational instability. Disk regions that cool below $T_p$ will cease to transport mass radially and dissipate energy. As such passive regions prevail all the way to the inner regions, the disk becomes passive, the accretion rate and disk torques turn off resulting in a sharp drop in  luminosity $L_x$  and in $\dot{P}$. Powered only by the cooling luminosity, such sources become unobservably faint, leading to the observed period clustering as the range of terminal periods at the time the fallback disk becomes passive. Inutsuka \& Sano (2005) found that viscosity generation is still possible at temperatures as low as 300 K, so $T_p$ should be less than 300 K. Our model evolutionary calculations yield  $T_p \sim$ 100 K - 200 K. There is an alternative evolutionary path that leads to period clustering in sources starting life with low initial disk mass. As the mass depletes, the torques drop down to very low levels so that the source settles asymptotically to $P \sim 10 s$ at an age $\sim 10^5 \,yrs$. XDINs may have evolved in this way.  

In all independent model applications to SGRs and AXPs, transient AXPs, XDINs and some radio pulsars, evolutionary models producing the present observed properties of the diverse sources work with the same torque model and require the same ranges for the physical parameters $C$, $\alpha_{hot}$ and $\alpha_{cold}$ (\cite[Ertan et al. 2007]{Ertanetal07}, \cite[Ertan \& Erkut 2008]{ErtanErkut08}, \cite[Alpar et al. 2011]{AEC11}, \cite[\c{C}al\i\c{s}kan et al. 2012]{Caletal12}, \cite[Ertan et al. 2012]{Ertanetal12}, \cite[Benli et al. 2012]{Benlietal12}). In this talk we discuss three sources, the AXP 4U 0142+61, the SGR 0418+5729 and the radio pulsar PSR J1734-3333 in the context of the fallback disk model.

\section{A Disk or Two Observed}

Mid-infrared band {\em Spitzer} observations of a disk around the AXP 4U 0142+61 were reported by \cite[Wang, Chakrabarty \& Kaplan(2006)]{WCK06}. This is clearly a fallback disk left over from the formation of the neutron star since, like all other AXPs (and other young neutron stars under discussion here), there is no evidence that 4U 0142+61 is in a binary. \cite[Wang, Chakrabarty \& Kaplan(2006)]{WCK06} interpreted this as a passive dust disk {\em outside the magnetosphere}, heated by X-ray irradiation and emitting solely in the mid infrared. Kern \& Martin (2002) had  observed optical pulses from 4U 0142+61 at the rotation period $P = 8.7\,s$. As the 27\% pulsed fraction in the optical is much larger than the pulsed fraction in X-rays, the optical pulsations cannot be due to disk reprocessing of the X-ray pulses in the disk, they must be of magnetospheric origin. \cite[Wang, Chakrabarty \& Kaplan(2006)]{WCK06} assumed that a magnetosphere with a disk inside cannot produce optical pulses. Hence they concluded that the disk they had observed was outside the magnetosphere, and emitted only the mid infrared radiation they had observed, and the optical and near infrared radiation, both pulsed and DC, came from a disk free magnetosphere.  However, as shown by \cite[Ertan \& Cheng (2004)]{ErtanCheng04}, based on the disk-dynamo model of \cite[Cheng \& Ruderman (1991)]{CR91}, the pulsed optical and near infrared emission {\em can} be produced from outer gaps in the pulsar magnetosphere operating with a disk inside. As for the DC emission, the combined optical, near and mid infrared observations can be fit with an irradiated gas disk model (\cite[Ertan et al. 2007]{Ertanetal07}). The best fits of Ertan et al. (2007) to the entire broad band DC data set from the optical to mid infrared yield the reddening $A_V = 3.5$, in good agreement with the $A_V \cong 3.5 \pm 0.4$ range indicated by independent observations \cite[Durant \& van Kerkwijk (2006)]{DvK06}. Another AXP, 1E 2259+586 has also been detected in the near infrared (Hulleman et al. 2001) and in the mid infrared with {\em Spitzer} (Kaplan et al. 2009). These detections are consistent with a disk source. A detection in the optical has not yet been made. 

\section{SGR 0418+5729: A Bursting Magnetar with Low Magnetic Dipole Field }

The SGR 0418+5729 with period $P = 9.1 s$ was found to have an exceptionally low spindown rate in comparison with other SGRs and AXPs. The upper limit 
$\dot{P} < 6 \times 10^{-15} s\; s^{-1}$ reported by \cite[Rea et al. (2010)]{Reaetal10} implies an  inferred dipole field $B_{dipole} < 7.5 \times 10^{12}\; G$ and characteristic age $\tau >  2.4  \times 10^{7}\; yrs$  if this source is assumed to be rotating in vacuum. A more recent, $3.5 \sigma$ measurement of $\dot{P} = 4 \times 10^{-15} s\; s^{-1}$ implies $B_{dipole} \cong 6 \times 10^{12}\; G$ and $\tau \cong  3.6 \times 10^{7}\; yrs$  \cite[(Rea et al. 2012b)]{Reaetal12b}. Within the isolated magnetar model this makes SGR 0418+5729 exceptional: in all other properties it is a typical SGR, while it differs from the class by orders of magnitude in its inferred dipole field. If assumed to be an isolated rotating dipole in vacuum, the position of SGR 0418+5729 in the $P\dot{P}$ diagram, well below the so-called death valley, makes it exceptional in comparison with the rotation powered pulsars: there is only one other source, the radio pulsar PSR J2144-3933 that is located at a similar outlying location in the $P\dot{P}$ diagram. Explanations within the magnetar model require substantial field decay and a relatively old age for this source (\cite[Turolla et al. 2011]{Turetal11}). 

The fallback disk model can explain all properties of this source in terms of interactive evolution of the neutron star with a fallback disk, employing the same fallback disk physics as previously applied to diverse other sources. The period, period derivative and X-ray luminosity of SGR 0418+5729 can be simultaneously reproduced. The dipole field strength indicated by our results is $1 - 2 \times 10^{12}\; G$ on the pole of the neutron star. The age of the star is $\sim 1 -  2 \times 10^{5}\; yrs$. The results are given in Fig.\,\ref{fig1}. 

There is now another SGR, Swift J1822.3-1606, with the second lowest measured $\dot{P} \cong 8.3 \times 10^{-14} s\; s^{-1}$ implying $B_{dipole} \cong 2.6 \times 10^{13}\; G$  (\cite[Rea et al. 2012a]{Reaetal12a}) or $\dot{P} \sim 3 \times 10^{-13} s\; s^{-1}$ implying $B_{dipole} \cong 5 \times 10^{13}\; G$ (\cite[Scholz et al. 2012]{Scholzetal12}). We are currently working on evolutionary models with a fallback disk for this source. Preliminary estimates for the surface dipole field give $B_{dipole} \sim 10^{12}\; G$. 

\section{PSR J1734-3333: A ``High-B" Radio Pulsar with a Peculiar Braking Index} 

The radio pulsar PSR J1734-3333, with period $P = 1.17 s $ and period derivative $\dot{P} = 2.28 \times 10^{-12} s\; s^{-1}$ has an inferred surface dipole field $B_{dipole} \cong 5.2 \times 10^{13}\; G$, reaching the range of inferred dipole fields for some AXPs. This pulsar is among the radio pulsars with the largest inferred dipole magnetic fields. Recent measurements showed that its period derivative $\dot{P}$ is increasing with time. The period second derivative was reported to have the value $\ddot{P} = 5.0(8) \times 10^{-24}\,s\;s^{-2}$ (\cite[Espinoza et al. 2011]{Espinoza11}). This corresponds to a braking index $n = 0.9 \pm 0.2$. Referring to the pulsar spindown relation $\dot{\Omega} = k\Omega^{n}$, equivalently expressed as $\dot{P} = K P^{2-n}$, where $k$ and $K$ depend on the magnetic moment and the  moment of inertia, a source with braking index $n$ will be evolving across the $P\dot{P}$ diagram on a line with slope $2-n$. For a dipole rotating in vacuum, $n = 3$. Most pulsars with measured braking indices have values of $n$ somewhat less than 3. With  $ n \cong 0.9$ PSR J1734-3333 is the exceptional outlier. To explain this braking index in terms of an isolated dipole would require either that the component of the dipole moment perpendicular to the rotation axis is increasing, i.e. the dipole moment is counter-aligning, shifting with respect to the neutron star, or the dipole component of the magnetic field, or perhaps the entire magnetic field is growing, on the $\sim 10^4~yrs$ characteristic spindown timescale of the pulsar. 

\begin{figure}[t]
\begin{minipage}[b]{0.5\linewidth}
\centering
\includegraphics[width=4.in,angle=270]{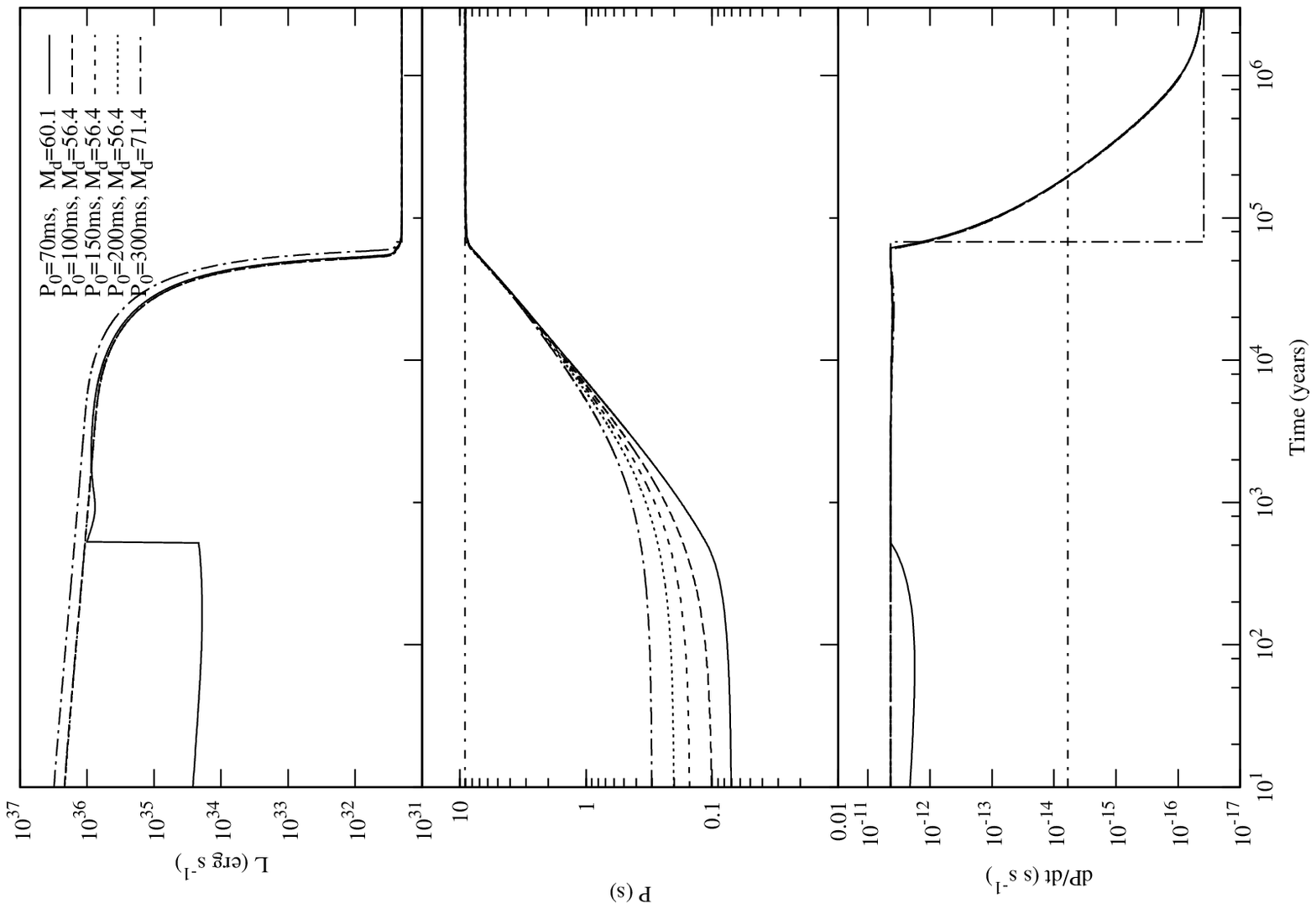}
\caption{$L_{\mathrm{x}}$, $P$ and $\dot{P}$ evolution of SGR 0418+5729: Figure 2 from \cite[Alpar, Ertan \& \c{C}al\i\c{s}kan (2011)]{AEC11}}
\label{fig1}
\end{minipage}
\hspace{0.5cm}
\begin{minipage}[b]{0.45\linewidth}
\includegraphics[width=4.in, angle=270]{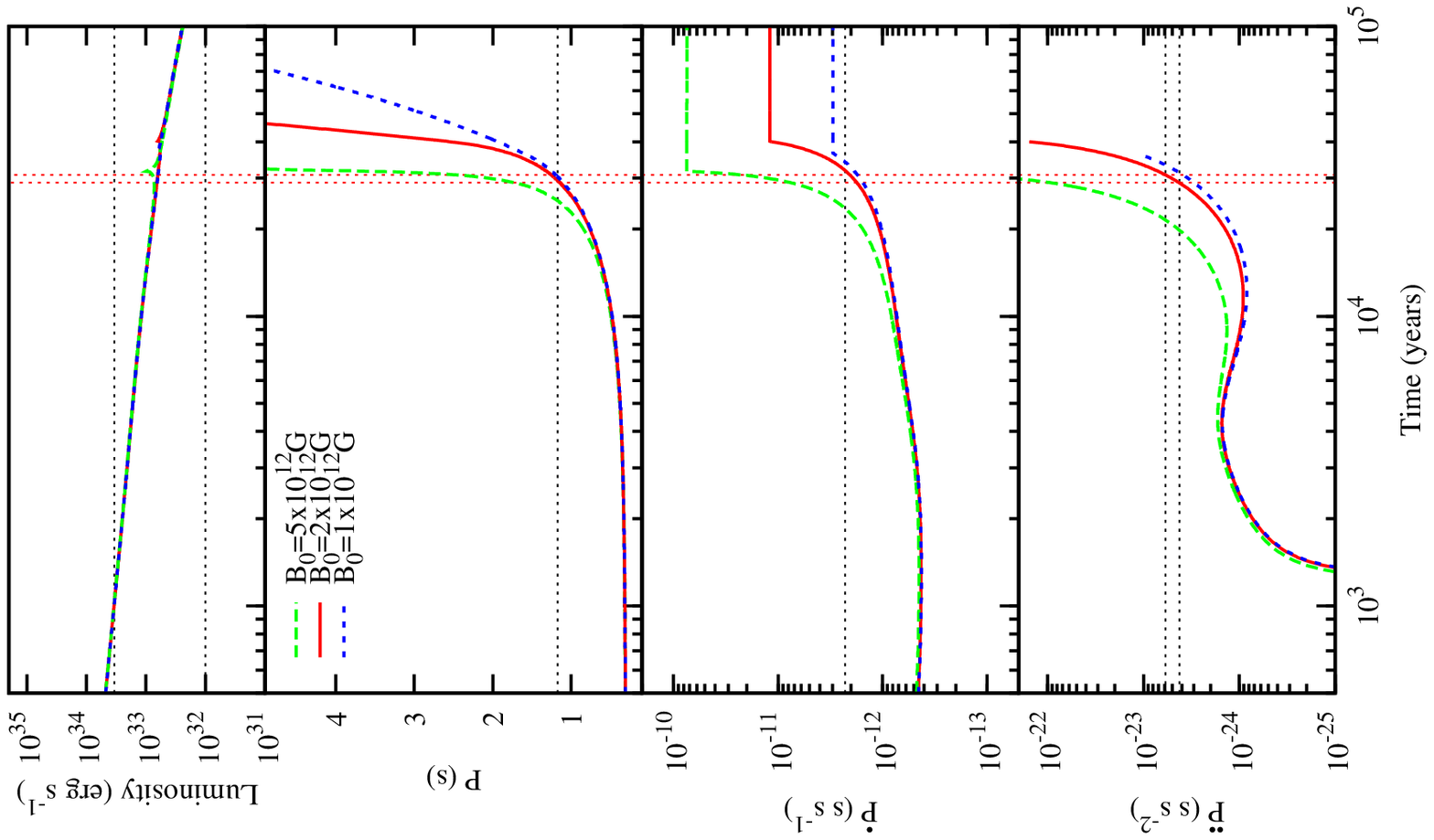}
\caption{$L_{\mathrm{x}}$, $P$, $\dot{P}$ and $\ddot{P}$ evolution of PSR J1734-3333: Figure 1 from \cite[\c{C}al\i\c{s}kan et al. (2012)]{Caletal12}}
\label{fig2}
\end{minipage}
\end{figure}

For neutron stars evolving with fallback disks, increasing $\dot{P}$ is expected in certain phases of the evolution. PSR J1734-3333 provides a more stringent challenge for the fallback disk scenario, compared to earlier applications of the model to other sources, as it is the first source with an observed $\ddot{P}$, in addition to the period, period derivative and luminosity, to be obtained at the present age. Using the same model as employed earlier to explain the evolution of anomalous X-ray pulsars and soft gamma-ray repeaters, we find that the period, the first and second period derivatives and the X-ray luminosity of this source can simultaneously acquire the observed values at an age of $\sim 3 \times 10^4\, yrs$ (\cite[\c{C}al\i\c{s}kan et al. 2012]{Caletal12}). The required strength of the dipole field that can produce the source properties is, once again, in the range of $ B_{dipole} \sim 1-3 \times 10^{12}\; G$. When the model source reaches the current state properties of PSR J1734-3333, accretion has not started yet, since the inner disk is not able to penetrate the light cylinder in this phase, allowing the source to operate as a regular radio pulsar. Such sources will have properties similar to the X-ray dim isolated neutron stars or transient AXPs at a later epoch of weak accretion from the diminished fallback disk. Our results are shown in Fig.\,\ref{fig2}.

\section{Summary and Discussion}

There is now strong evidence from different directions, that the presence and properties of a fallback disk around a young neutron star may determine its evolution and explain the various classes of young neutron stars. First, a fallback disk has been observed around the AXP 4U 0142+61. An active gaseous fallback disk fits the data from multiple bands. Secondly, models of fallback disks in interaction with the neutron star give successful evolutionary scenarios that can reproduce the physical parameters of persistent and transient AXPs, SGRs, XDINs and some radio pulsars with exceptional properties. All of these applications indicate surface dipole field strengths in the range $B_{dipole} \sim 10^{12} - 10^{13}\; G$. Two recent examples were discussed here. SGR 0418+5729 shows that the dipole field, even as inferred for dipole spindown in vacuum, can be indeed significantly lower than the total field required to power magnetar bursts. PSR J1734-3333, with its similarities and differences from the magnetars, can be understood in terms of evolution with a fallback disk. If seen as isolated pulsars, SGR 0418+5729 must have experienced rapid decay of the dipole field, while PSR J1734-3333 must have a rapidly growing dipole field at present. If further the total magnetic field is in the magnetar range, such ongoing growth of the dipole field is difficult to understand, unless, perhaps, the dipole field was buried and is now reemerging (\cite[Pons et al. 2012]{Ponsetal12}). 

The observations show that the dipole field does not have to be of the same magnitude as the total field, as we have been emphasizing for some time. The possibility of sustained presence of matter with angular momentum (fallback disks) is also supported by the discovery of a disk around at least one isolated neutron star, 4U 0142+61. The two sources, SGR  0418 + 5729 and PSR J1734-3333, like other young neutron stars, are thus linked in a broader picture where, from a single population as regards dipole moments and initial periods, an additional initial condition, the possible presence and properties of a fallback disk, determines the evolutionary pathways leading to the diverse classes of young neutron stars.

\end{document}